\documentclass[aps,prd,amsmath,preprint]{revtex4}

\usepackage[utf8]{inputenc}
\usepackage{amsmath,amssymb}
\usepackage{bm}
\usepackage{amscd}	
\usepackage{graphicx}
\usepackage{dcolumn}
\usepackage{url}
\usepackage{color}
\usepackage[colorlinks=true,linkcolor=blue]{hyperref}

\preprint{JLAB-THY-14-1856}

\begin{document}

\title{Constraints on spin-dependent parton distributions at large $x$ \\
	from global QCD analysis}
\author{P.~Jimenez-Delgado, H.~Avakian,	W.~Melnitchouk}
\affiliation{\mbox{Jefferson Lab, Newport News, Virginia 23606, USA} \\
	{\bf Jefferson Lab Angular Momentum (JAM) Collaboration}\\
}

\date{\today\\}

\begin{abstract}
We investigate the behavior of spin-dependent parton distribution
functions (PDFs) at large parton momentum fractions $x$ in the
context of global QCD analysis.
We explore the constraints from existing deep-inelastic scattering
data, and from theoretical expectations for the leading $x \to 1$
behavior based on hard gluon exchange in perturbative QCD.
Systematic uncertainties from the dependence of the PDFs on the
choice of parametrization are studied by considering functional
forms motivated by orbital angular momentum arguments.
Finally, we quantify the reduction in the PDF uncertainties that may
be expected from future high-$x$ data from Jefferson Lab at 12~GeV.
\end{abstract}

\maketitle

\section{Introduction}
\label{sec:intro}

Recently a new global next-to-leading order (NLO) analysis
\cite{JAM-DIS} of spin-dependent parton distribution functions (PDFs)
was performed by the JAM (Jefferson Lab Angular Momentum) Collaboration
\cite{JAMweb}, in which particular attention was paid to the valence
quark-dominated region at high parton momentum fractions $x$ and low
four-momentum transfers $Q^2$.
This region requires careful treatment of the potentially important
$1/Q^2$ power corrections associated with target mass and higher
twist contributions to the inelastic cross sections, as well as
nuclear smearing effects when using deuterium and $^3$He data.
The analysis \cite{JAM-DIS} indeed found significant effects on
the polarized leading twist PDFs when twist-3 and twist-4 power
corrections in both the spin-dependent $g_1$ and $g_2$ structure
functions were taken into account.
In particular, the $\Delta d^+$ distribution (defined as
$\Delta d^+ \equiv \Delta d + \Delta \bar d$) was found to have a
significantly larger magnitude at $x \gtrsim 0.2$ than in previous
global analyses, driven partly by a large and positive twist-4
correction to the neutron $g_1$ structure function.

Analyses such as those in Ref.~\cite{JAM-DIS} that systematically
incorporate subleading effects in an effort to accommodate data
over a broad range of kinematics can therefore provide a more
solid basis for extracting reliable information on PDFs and their
uncertainties, especially in regions such as at large $x$ where
data are relatively scarce \cite{PJD13}.  In fact, the absence of
high-precision polarization data at high $x$, particularly for
the neutron (or $^3$He), has meant that spin-dependent PDFs are
essentially unconstrained for $x \gtrsim 0.5-0.6$.
This is rather unfortunate, given that polarized PDFs, and ratios of
polarized to unpolarized PDFs, are quite sensitive to the details of
nonperturbative quark-gluon dynamics in the nucleon at high $x$,
with theoretical predictions differing in some cases even in sign
\cite{MT96}.

In the simplest quark models, for example, spin-flavor symmetry
implies constant ratios of PDFs,
$\Delta u/u = 2/3$,
$\Delta d/d = -1/3$ and
$\Delta d / \Delta u = - 4$.
Symmetry breaking effects, which typically generate a larger
energy for axial vector spectator diquark configurations compared
to scalar diquarks, generally raise the $\Delta u/u$ ratio to unity
in the $x \to 1$ limit, while keeping $\Delta d/d$ unchanged from
the SU(6) value \cite{Feynman72, Close73, Close88, Isgur99}.
Calculations of one-gluon exchange in perturbative QCD (pQCD),
on the other hand, predict that \cite{Farrar75}
\begin{eqnarray}
\frac{\Delta q(x)}{q(x)} &\to& 1 \hspace*{0.5cm} {\rm as}\ \ x \to 1,
\label{eq:x1limit}
\end{eqnarray}
where $q(x)$ is the spin-averaged distribution, for all quark
flavors $q$.
Similar expectations arise also from arguments based on local
quark-hadron duality \cite{BG70, WM01, MEK05, Close03}.

While most global PDF analyses do not include conditions such as in
Eq.~(\ref{eq:x1limit}) in order to avoid introducing theoretical bias
into the PDF extraction, Brodsky, Burkardt and Schmidt \cite{BBS95}
proposed a simple parametrization of PDFs in which the large-$x$
constraints of Eq.~(\ref{eq:x1limit}) were built in.
Only a limited set of data was analysed in Ref.~\cite{BBS95},
however, although a subsequent global analysis utilizing the pQCD
expectations was performed by Leader, Sidorov and Stamenov (LSS)
\cite{LSS98}.
This found that a reasonably good fit to the available data was
indeed possible, with the feature of a steep rise in the
$\Delta d/d$ ratio at intermediate values of $x$.
Later high-precision data from the E99-117 experimental at
Jefferson Lab \cite{E99-117} observed the first evidence of a turn
over in the $A_1^n$ neutron polarization asymmetry from negative
to positive values, although at somewhat larger values of $x$
($x \sim 0.5-0.6$) than those in the fit of Ref.~\cite{LSS98}.

In a more recent analysis, Avakian {\it et al.} \cite{Avakian07}
showed that inclusion of $L_z=1$ components in the lowest three-quark
Fock state of the nucleon, in addition to the usual $L_z=0$
configurations, can generate additional terms that behave as
$(1-x)^5 \log^2(1-x)$ at large $x$, which can play an important
role numerically.
%
%
Generalizing the pQCD-inspired parametrization from Ref.~\cite{BBS95}
to include the subleading $\log^2(1-x)$ terms, Avakian {\it et al.}
showed that the large-$x$ asymmetry data could be well fitted while
preserving the asymptotic constraints of Eq.~(\ref{eq:x1limit}).
In particular, the new $\Delta d/d$ ratio was found to remain
negative until $x \approx 0.75$, as suggested by the E99-117 data
\cite{E99-117}, before rising towards unity at higher $x$ values.

While the analysis of Ref.~\cite{Avakian07} showed the potential
of high-$x$ data to reveal information about the orbital angular
momentum (OAM) of quarks in the nucleon, it was not based on a
comprehensive global analysis of all available data.  The goal of
the present work is to examine the behavior of spin-dependent PDFs
in the $x \sim 1$ region in the context of a global QCD analysis,
including the effects of the $x \to 1$ constraints in
Eq.~(\ref{eq:x1limit}) and of the $\log^2(1-x)$ terms inspired
by pQCD.

We begin our discussion in Sec.~\ref{sec:fits} by summarizing the
recent global analysis \cite{JAM-DIS} from the JAM Collobaration,
which we use as the baseline fit for our large-$x$ studies.
To explore the dependence on the choice of parametrization and allow
for more direct connection with quark orbital angular momentum,
we also consider a simplified functional form which uses a smaller
number of parameters.
The effects on the fits of additional terms in the PDF
parametrizations induced by nonzero orbital angular momentum
are investigated, together with the impact on the $\Delta u$
and $\Delta d$ PDFs from imposing constraints for the $x \to 1$
behavior from perturbative QCD.
In Sec.~\ref{sec:12gev} we repeat the global analysis using in
addition pseudodata generated at the kinematics of future data
from several experiments planned at the 12~GeV energy upgraded
Jefferson Lab, and quantify the resulting reduction in the PDF
errors at high $x$.
Finally in Sec.~\ref{sec:conc} we draw some conclusions of the
present analysis and outline steps for future work.

\section{Parton distributions at large \large $x$ \normalsize}
\label{sec:fits}

For our exploration of the large-$x$ region we use as a baseline the
PDFs from the JAM analysis in Ref.~\cite{JAM-DIS}.  The JAM PDFs were
obtained from a global NLO fit to all available data on inclusive
polarized deep-inelastic scattering asymmetries for
\mbox{$Q^2 > Q_0^2 = 1$~GeV$^2$} and $W > 1.87$~GeV.
Inclusion of low-$Q^2$ and low-$W$ data necessitated a careful
treatment of the subleading $1/Q^2$ contributions, to both the $g_1$
and $g_2$ structure functions, from target mass and higher twist
corrections, as well as nuclear smearing effects for deuterium
and $^3$He data.
By fitting directly the longitudinal and transverse polarization
asymmetries, where available, one avoids introducing biases that would
otherwise arise in fits to the spin-dependent structure functions,
which are often extracted from the experimental asymmetries under
different assumptions for the spin-averaged structure functions.

\begin{figure}[h]
\includegraphics[width=8cm]{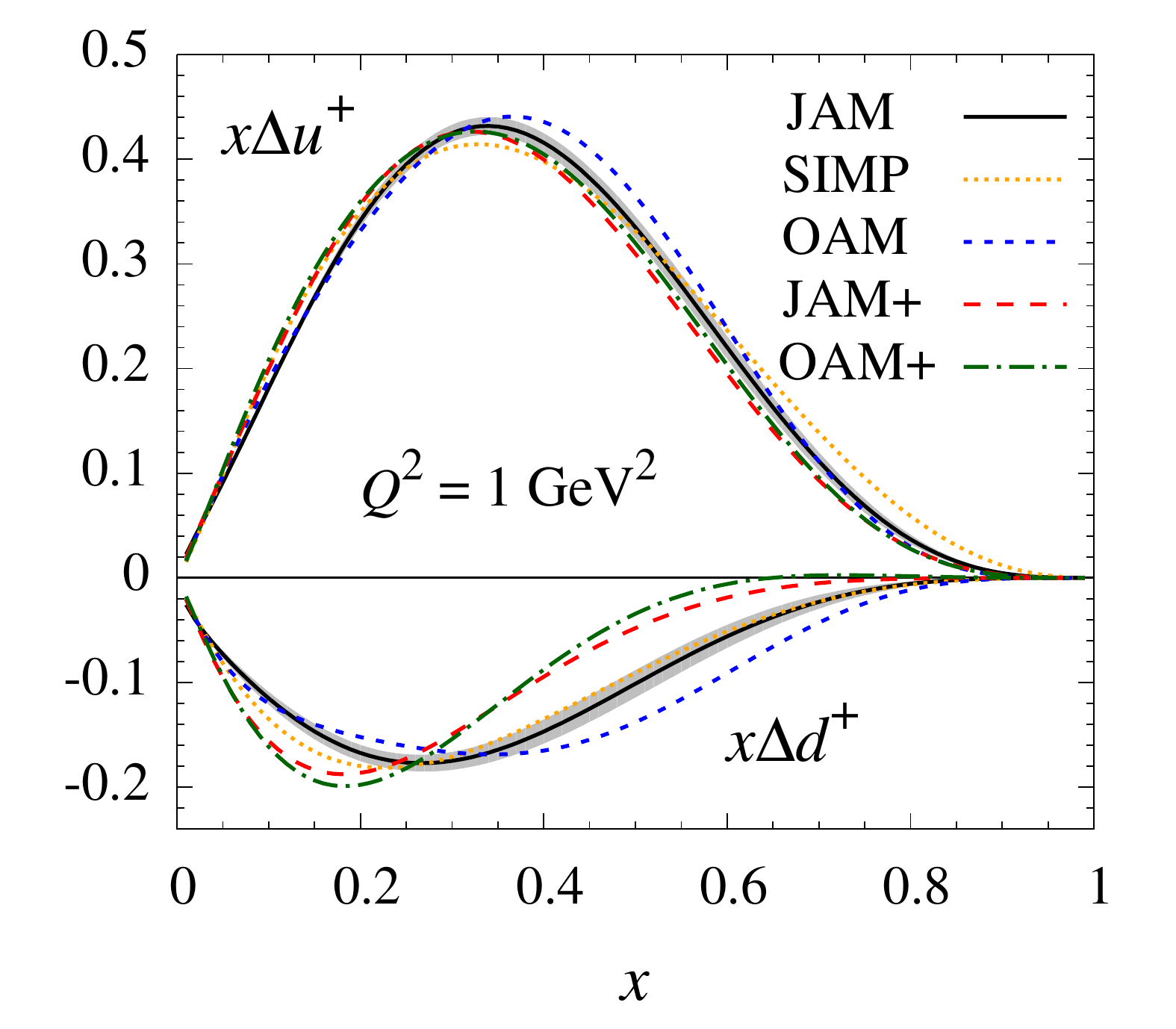}
\includegraphics[width=8cm]{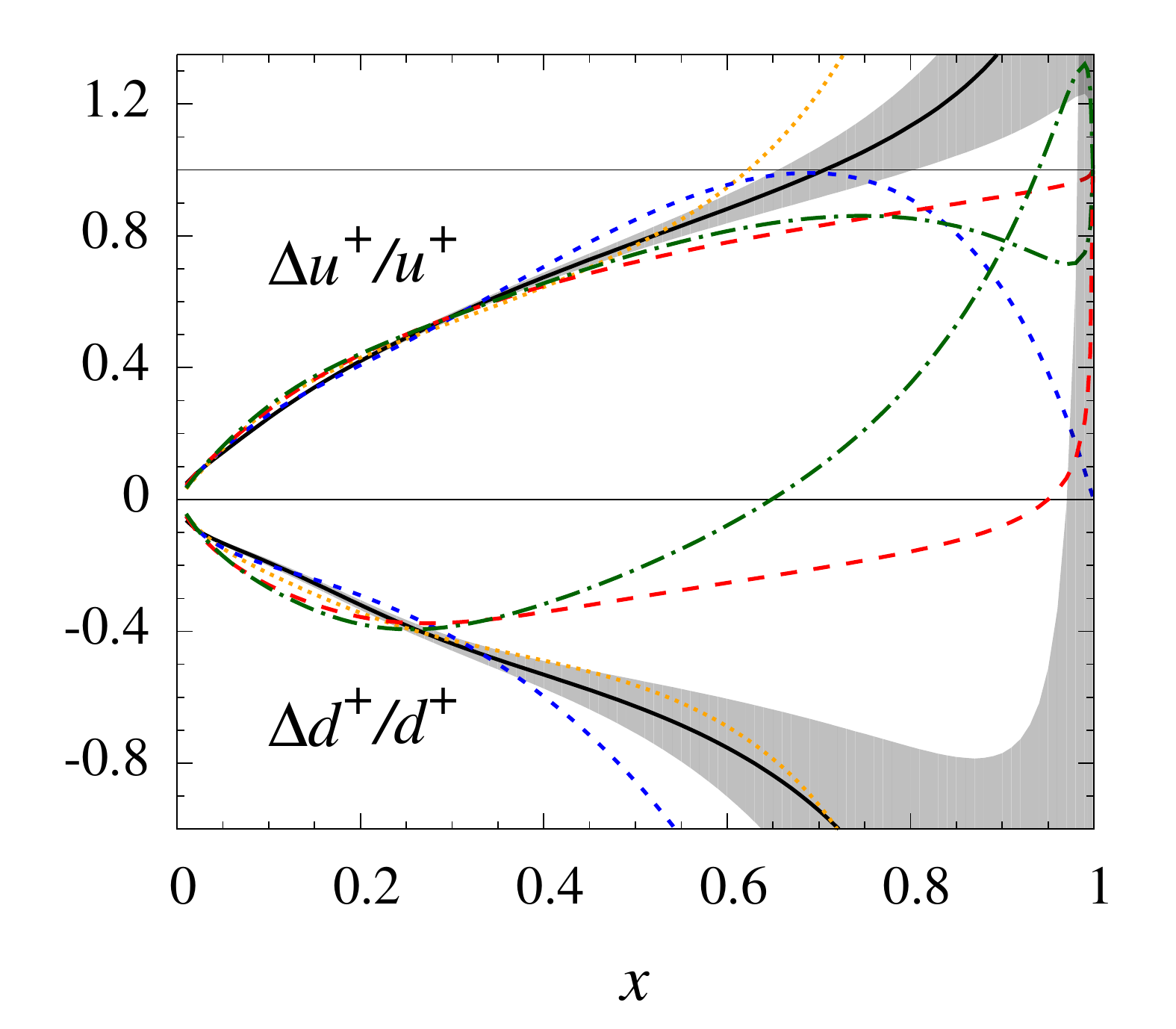}
\caption{Spin-dependent parton distributions $\Delta u^+$
	and $\Delta d^+$ {\bf (left)} and their ratios
	$\Delta u^+/u^+$ and $\Delta d^+/d^+$ to the spin-averaged
	distributions {\bf (right)} at a scale of $Q^2 = 1$~GeV$^2$.
	The distributions from the JAM analysis \cite{JAM-DIS}
	(black solid curves) are compared with those using a more
	basic parametrization (orange dotted curves), the ``OAM''
	parametrization inspired by nonzero orbital angular
	momentum considerations (blue short-dashed curve),
	and the ``JAM+'' (red long-dashed curves) and ``OAM+''
	(green dot-dashed curves) fits which include the
	$x \to 1$ constraint from Eq.~(\ref{eq:x1limit}).}
\label{fig:qplus}
\end{figure}

A standard parametrization was used for the polarized quark,
antiquark and gluon distributions in terms of four parameters
plus an overall normalization,
\begin{eqnarray}
x \Delta q^+(x,Q_0^2)
&=& N\, x^\alpha (1-x)^\beta\, (1 + \epsilon \sqrt{x} + \eta\, x),
\label{eq:JAM}
\end{eqnarray}
where $\Delta q^+ \equiv \Delta q + \Delta\bar{q}$, at the input
scale $Q_0^2$.  At large values of $x$ the antiquark and gluon PDFs
play a negligible role, so that the inclusive DIS data alone are
sufficient to determine the $\Delta u^+$ and $\Delta d^+$
distributions individually.
A total of over 1,000 data points were used in the analysis, giving
an overall $\chi^2$ per data point of 0.98 (see Table~\ref{tab:chi2}).
The resulting $\Delta u^+$ and $\Delta d^+$ distributions are
illustrated in Fig.~\ref{fig:qplus}, together with the ratios to
the spin-averaged PDFs, $\Delta u^+/u^+$ and $\Delta d^+/d^+$,
at $Q^2=1$~GeV$^2$.

\begin{table}[ht]
\begin{center}
\caption{$\chi^2$ values per number of data points $N_{\rm dat}$
	for the various fits discussed in this analysis,
	including the JAM, SIMP	and OAM	fits, with or without
	the $x \to 1$ constraint in Eq.~(\ref{eq:x1limit}),
	and including leading twist (LT) contributions only.
	For the JAM fit we also considered the cases where the
	$\Delta d^+$ PDF was forced to cross zero at $x=0.5$
	and $x=0.75$.\\}
\begin{tabular}[c]{l|ccc|ccc}
\hline
\hspace{1.3cm}$\chi^2/N_{\rm dat}$
	&\ \ JAM\  &\ \ OAM\  &\ \ SIMP\ \
	&\ JAM(LT) &\ OAM(LT) &\ SIMP(LT)\	\\ \hline
no $x \to 1$ constraint
	&\ \ 0.98  &\ \ 0.98  &\ 1.02 
	&  1.07    &\   1.09  & 1.12		\\
with $x \to 1$ constraint\ \
	&\ \ 1.01  &\ \ 1.02  &\ ---
	&  1.11    &\ 1.16    & --- 		\\
$\Delta d^+$ crossing at $x=0.75$
	&\ \ 1.02  &\ \ ---   &\ ---
	&  ---     &\ ---     & ---		\\
$\Delta d^+$ crossing at $x=0.5$
	&\ \ 1.06  &\ \ ---   &\ ---
	&  ---     &\ ---     & --- 		\\ \hline
\end{tabular}
\label{tab:chi2}
\end{center}
\end{table}

Unlike the $Q^2$ dependence of PDFs, which is determined by the $Q^2$
evolution equations to a given order in the strong coupling $\alpha_s$
\cite{Weigl96}, the $x$ dependence of PDFs is generally not accessible
directly from pQCD calculations.  An exception is the kinematic region
at large $x$, where hard gluon exchange between the quarks in the
leading three-quark Fock state component of the nucleon can be used
to determine the dominant contributions to the $x$ dependence of the
PDFs in the $x \to 1$ limit \cite{Farrar75}.
Typically one finds that the quark PDFs in the nucleon behave as
$\sim (1-x)^{2n_s-1}$, where $n_s$ is the minimum number of partons
that are spectators to the hard collision \cite{Farrar75, BBS95},
so that for $n_s=2$ the leading exponent is equal to 3.
More generally, the exponent on $(1-x)$ also depends logarithmically
on $Q^2$ \cite{Carlson94}, although the scale from which this should
evolve is {\it a priori} unknown.  Nevertheless, global PDF fits do
find parameters $\beta$ in Eq.~(\ref{eq:JAM}) that are close to the
pQCD (or quark ``counting rule'') predictions; for the JAM fit, for
instance, one has $\beta_u=3.3$ and $\beta_d=4.0$ for $\Delta u^+$
and $\Delta d^+$, respectively, at the input scale $Q_0^2$.

Of course, the additional polynomial terms in (\ref{eq:JAM}) with
coefficients $\epsilon$ and $\eta$ obscure the direct connection
between the $x$ dependence of the fitted distributions and the
predicted pQCD behavior.  To make the connection more explicit,
we consider a fit based on a simplified functional form, with
parameters $\epsilon$ and $\eta$ in Eq.~(\ref{eq:JAM}) set to zero.
The resulting fit, labeled ``SIMP'' in Fig.~\ref{fig:qplus}, gives
similar $\Delta u^+$ and $\Delta d^+$ distributions to those from
the full JAM analysis, albeit with a slightly larger overall
$\chi^2$ value.  The leading $(1-x)$ exponents in this case are
reduced slightly to $\beta_u=2.5$ and $\beta_d=3.4$.

As well as specifying the leading $x \to 1$ behavior of the PDFs,
the pQCD counting rules also predict that the dominant contribution
to the cross section, in a reference frame where the nucleon is
moving fast along the $z$-axis, is from scattering off quarks
with the {\it same} helicity as that of the nucleon.
This implies that asymptotically the helicity-aligned distributions
dominate both the unpolarized and polarized PDFs, so that the ratio
$\Delta q^+/q^+ \to 1$ as $x \to 1$ for all quark flavors $q$, as
in Eq.~(\ref{eq:x1limit}).  In this scenario the $A_1$ polarization
asymmetries in DIS are therefore expected to approach unity for
both the proton and neutron.

Unfortunately, current data cannot definitively confirm the pQCD
expectations.  While the proton $A_1^p$ asymmetries, which have
been measured to $x \approx 0.7$ in the DIS region, are consistent
with an approach towards unity in the $x \to 1$ limit, the neutron
(or $^3$He) data extend only to $x \approx 0.6$ and are generally
consistent with a zero or negative asymmetry.
The dearth of high-$x$ data means that the spin-dependent PDFs,
and particularly the $\Delta d^+$ distributions, are essentially
unconstrained above this region.  Consequently the spin-dependent
PDFs obtained from global analyses often violate the positivity
condition $|\Delta q(x)| \leq q(x)$ at large $x$ (although strictly
speaking these need not be satisfied beyond leading order). 
This can be seen in Fig.~\ref{fig:qplus} for both the JAM and
SIMP (and other) fits, and in general will be a feature of any
global fit which does not {\it a priori} impose the positivity
constraint (or else fit the individual helicity-aligned and
helicity-antialigned distributions separately).
Note that in the present work the unpolarized fits are based
on the recent NLO global analysis from Ref.~\cite{JR14}.
Ideally one should perform a global fit of polarized and unpolarized
data simultaneously and extract the helicity-aligned and antialigned
distributions.  This will be particularly important in future analyses
that include semi-inclusive data sets, where transverse momentum
distributions, which are expected to be different for different
helicity states \cite{Musch11}, will play a more important role.

\begin{figure}[t]
\includegraphics[width=8cm]{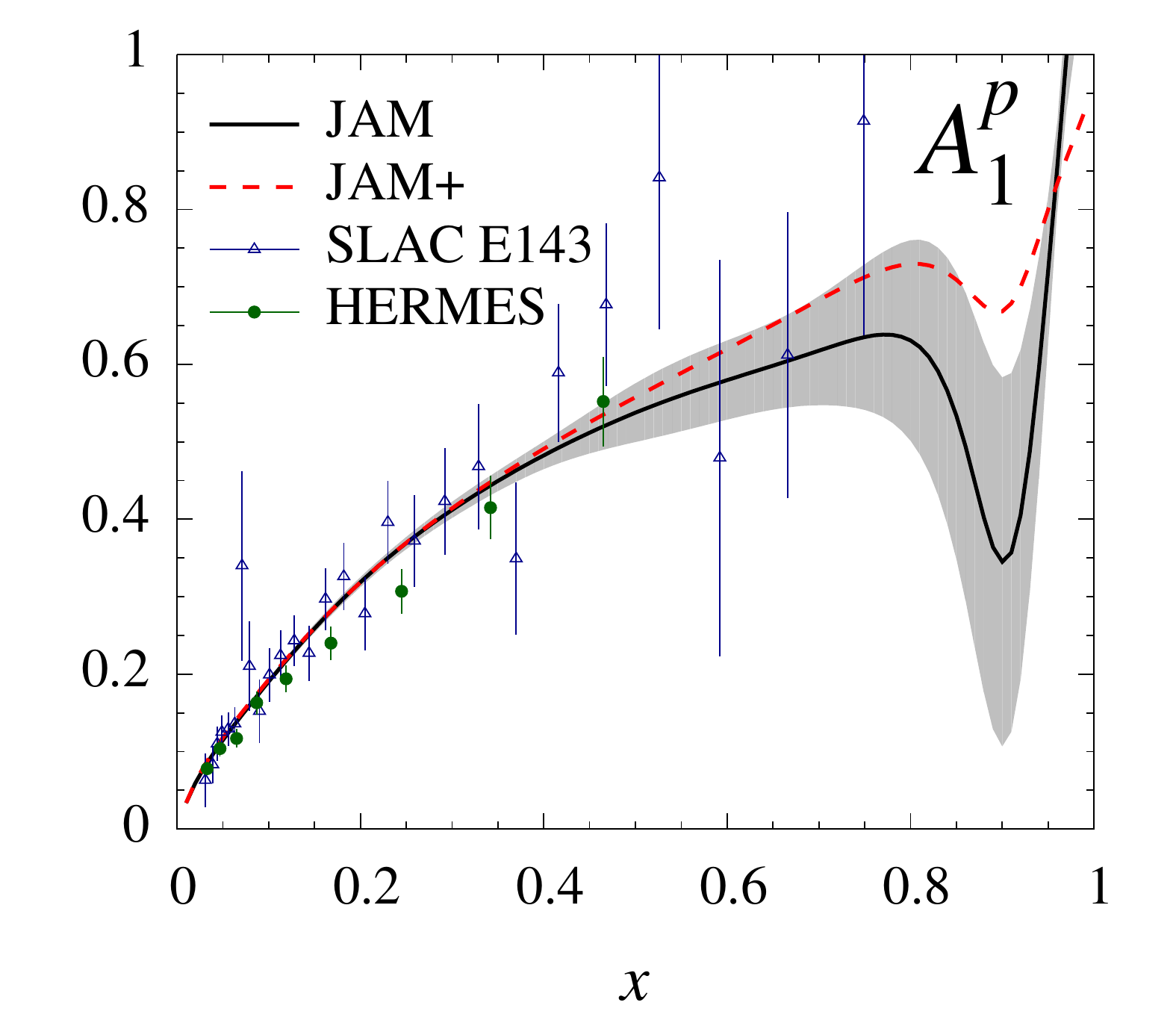}
\includegraphics[width=8cm]{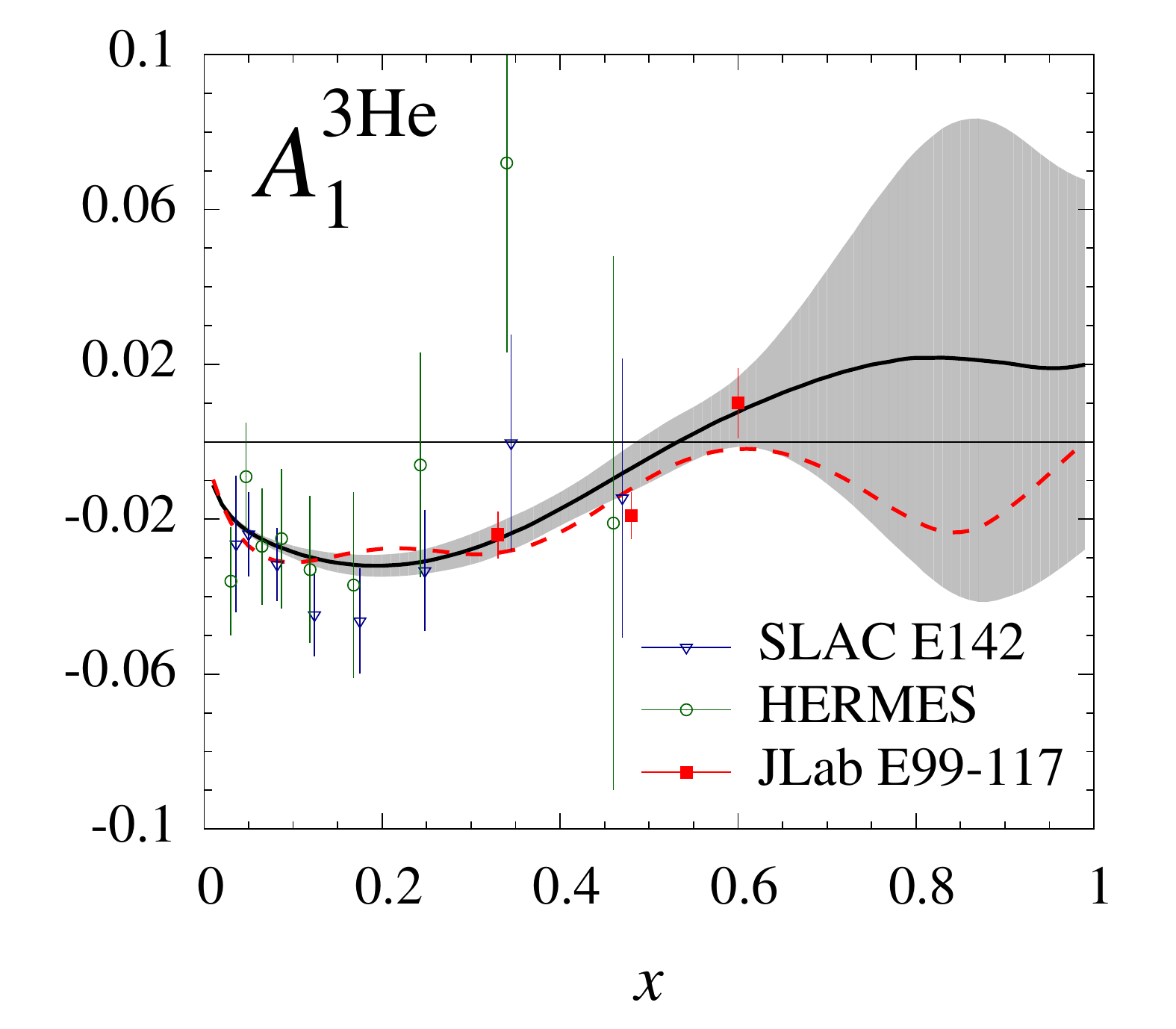}
\caption{Comparison of JAM (black solid curves and gray band)
	and JAM+ (red dashed curves) fits at $Q^2=3$~GeV$^2$
	with proton {\bf (left)} and $^3$He {\bf (right)} $A_1$
	polarization asymmetry data for similar $Q^2$ values.
	The proton data are from the SLAC E143 \cite{SLAC-E143}
	and HERMES \cite{HERMES-p} experiments, while the $^3$He
	data are from the SLAC E142 \cite{SLAC-E142}, HERMES
	\cite{HERMES-He} and Jefferson Lab E99-117 \cite{E99-117}
	experiments.}
\label{fig:data}
\end{figure}

To explore the effect of the $x \to 1$ constraints (\ref{eq:x1limit})
on PDFs in the context of the JAM global analysis, we consider a
modified fit in which the $\Delta u^+/u^+$ and $\Delta d^+/d^+$
ratios are both forced to unity at $x=1$.  The resulting fit,
denoted by ``JAM+'' in Fig.~\ref{fig:qplus} and Table~\ref{tab:chi2},
shows that one can indeed obtain a reasonable description of data,
consistent with the pQCD $x \to 1$ limit, with an overall
$\chi^2/N_{\rm dat} = 1.01$ that is only slightly larger than
for the unconstrained fit. 
This is confirmed also in Fig.~\ref{fig:data}, where the JAM and
JAM+ fits are compared with data on the $A_1$ asymmetries for the
proton and $^3$He from SLAC, HERMES, and Jefferson Lab.
The increase in $\chi^2$ is associated with the reduced magnitude
of $\Delta d^+$ in the intermediate-$x$ region, $x \gtrsim 0.3$,
which in order to maintain the normalizations required by the
triplet and octet axial vector charges \cite{JAM-DIS}, becomes
slightly more negative (with larger magnitude) at smaller $x$,
$x \lesssim 0.2$, where considerably more data exist.
The $\Delta u^+$ distribution, on the other hand, which was
already large and positive in the JAM fit, undergoes relatively
little change with the $x=1$ constraint.
Note that the constraint (\ref{eq:x1limit}) cannot be accommodated
by the SIMP parametrization, as without nonzero $\epsilon$ or
$\eta$ terms in Eq.~(\ref{eq:JAM}) the $\Delta d^+$ distribution
cannot change sign at any $x$.

Interestingly, the turn-over in $\Delta d^+$ from negative
to positive values occurs at relatively large values of $x$,
$x \approx 0.95$, which would be challenging to observe
experimentally.  This is significantly higher than the
turn-over found in the earlier LSS analysis \cite{LSS98} at
$x \sim 0.5$, which was subsequently found to be in conflict
with the neutron asymmetry data from the E99-117 experiment
at Jefferson Lab \cite{E99-117}.
Indeed, the existing data tend to disfavor fits with positive
$d$ quark polarization over the measured $x$ range.
We studied this by forcing a zero crossing in $\Delta d^+$ at
$x = x_0$, with the distribution becoming positive for $x > x_0$.
For the JAM fit with $x_0 = 0.75$ the $\chi^2/N_{\rm dat}$
increased slightly compared to the JAM+ fit, but the increase
was significantly greater, to $\chi^2/N_{\rm dat} = 1.06$,
when the crossing was set at a lower value, $x_0 = 0.5$.

Of course, the behavior of leading twist PDFs at large $x$ is also
influenced to some extent by the effect of higher twist corrections,
which become more important as $x \to 1$.  Using either the JAM
or SIMP parametrizations {\it without} including the $1/Q^2$
corrections generally results in a significantly worse fit,
with $\chi^2/N_{\rm dat}$ values increasing form $\approx 1$
to $\approx 1.1$ in Table~\ref{tab:chi2}, regardless of whether
the constraint (\ref{eq:x1limit}) is imposed or not.
This supports the findings of Ref.~\cite{JAM-DIS} that the $Q^2$
dependence of the data over the range considered here cannot be
accommodated by the parametric form in Eq.~(\ref{eq:JAM}) with
$Q^2$ corrections from $Q^2$ evolution only.

To address the problem of the rapid rise of $\Delta d^+$ at too low
values of $x$, Avakian {\it et al.} \cite{Avakian07} generalized the
pQCD calculations for the $x \to 1$ behavior of PDFs by considering
components of the lowest three-quark Fock-state wave function with
nonzero orbital angular momentum, $L_z=1$, in addition to the usual
$L_z=0$ configurations.  The $L_z=1$ contributions were found to
generate additional terms that behave as $\sim (1-x)^5 \log^2(1-x)$
at large $x$.  Although formally subleading in the $x \to 1$ limit
compared with the dominant $\sim (1-x)^3$ contributions expected
from the $L_z=0$ component, numerically the $\log$ terms can play
an important role.
In particular, Avakian {\it et al.} found that by using the
pQCD-inspired parametrization from Ref.~\cite{BBS95} supplemented
by the subleading $\log^2(1-x)$ terms, the large-$x$ asymmetry data
could be well fitted while preserving the asymptotic constraints
of Eq.~(\ref{eq:x1limit}).  Furthermore, the resulting
$\Delta d^+/d^+$ ratio remained negative until $x \approx 0.75$,
as suggested by the E99-117 data \cite{E99-117}, before rising
towards unity at higher $x$ values.

To explore the importance of the additional $\log^2(1-x)$ terms
in the context of a global QCD analysis of all data, at small
and high $x$, we use as a basis the simplified parametrization
with $\epsilon = 0 = \eta$ in Eq.~(\ref{eq:JAM}), together with
the $\log$ term inspired by the OAM arguments,
\begin{eqnarray}
x \Delta q^+
&=& N\,  x^\alpha    (1-x)^\beta\
 +\ N'\, x^{\alpha'} (1-x)^{\beta'} \log^2(1-x).
\label{eq:OAM}
\end{eqnarray}
with arbitrary relative normalization $N'$.
It is reasonable as a first approximation to assume that the
$x \to 0$ behavior of the OAM-inspired term is the same as the
standard term, $\alpha' = \alpha$.  To reduce the number of
parameters that can be reliably determined from the existing data,
we also fix $\beta'=5$ in accordance with the pQCD derivation
\cite{Avakian07}, even though the corresponding power of $(1-x)$
in the first term of Eq.~(\ref{eq:OAM}) remains a free parameter.

The resulting fit, denoted by ``OAM'' in Fig.~\ref{fig:qplus},
is of comparable quality to the JAM fit ($\chi^2/N_{\rm dat}=0.98$),
with similar $\Delta u^+$ and $\Delta d^+$ distributions at
moderate $x \lesssim 0.4$, but differing at higher $x$ values,
where there are no constraints from data.  If one includes in
addition the $x=1$ constraint from Eq.~(\ref{eq:x1limit}),
the effect on the new constrained fit, labelled ``OAM+'' in
Fig.~\ref{fig:qplus}, is again similar to that on the JAM+ fit.
Namely, the $\Delta d^+$ PDF is forced to become positive at
$x \approx 0.65$, and the reduced magntitude forces the distribution
at smaller $x$ values to become more negative in order to preserve
the sum rules.  The $\Delta u^+$ distribution remains relatively
unchanged, and the overall $\chi^2/N_{\rm dat} = 1.02$ is
comparable to that for JAM+.
%
%
The OAM and OAM+ fits with LT contributions only are once again
considerably worse than the full fits including higher twist
effects, indicating that the need for subleading corrections is
independent of the parametric form chosen for the LT component.

The results of the above fits suggest that with the additional
flexibility afforded by the $\log^2(1-x)$ terms in Eq.~(\ref{eq:OAM}),
the current data certainly {\it can} be accommodated with the
OAM-inspired parametrization.  On the other hand, the JAM and JAM+
fits based on the standard parametrization in Eq.~(\ref{eq:JAM}) give
perfectly good descriptions of the available data over the entire
range of kinematics, and {\it do not need} the introduction of the
additional $\log$ terms.  The constraint from Eq.~(\ref{eq:x1limit}),
when imposed on the standard PDFs, can be satisfied without
substantially modifying the distributions in the regions
constrained by data.
One should also caution, however, that the $\log^2(1-x)$ term in the
OAM-inspired parametrization (\ref{eq:OAM}) cannot at present be
directly related to the component of the nucleon's spin carried by
the quark orbital angular momentum \cite{Leader13}.  Its appearance
in the present analysis serves more to illustrate the possible role
played by OAM in understanding PDFs at large $x$, and to explore the
systematic uncertainties that may arise from different assumptions
about the functional forms used for the PDF parametrizations.
Fits including only terms with $L_z=0$ and $L_z=1$ \cite{Avakian07},
which can be interpreted in terms of relative contributions from
different orbital states, will be reported elsewhere \cite{JAM-future}.

A scenario in which one finds qualitatively different fits with
comparable $\chi^2$ values, or fits which differ by amounts that are
larger than the uncertainties from the propagation of experimental
errors, indicates a lack of information at large $x$, and an
underestimate of the systematic errors in this region.
In the absence of clearer theoretical constraints at $x \lesssim 1$,
the problem can be best addressed of course by the availability of
new data at higher $x$ values than are currently available, which
we discuss in the next section.

\section{Constraints from future data}
\label{sec:12gev}

Constraining the behavior of the polarization asymmetries $A_1$,
and consequently of the spin-dependent PDFs, in the limit as
$x \to 1$ is one of the featured goals of the experimental physics
program planned for the 12~GeV energy upgraded CEBAF accelerator
at Jefferson Lab.  Data from several experiments are expected to
be collected for values of $x$ as high as $\approx 0.8$ for DIS
kinematics \cite{12GeV}, and even higher $x$ in the nucleon
resonance region.  This should significantly reduce the PDF
uncertainties for $x \gtrsim 0.5$, especially for the $\Delta d^+$
distribution, which will be more strongly constrained by new data
on the $^3$He asymmetry.

\begin{figure}[t]
\includegraphics[width=8cm]{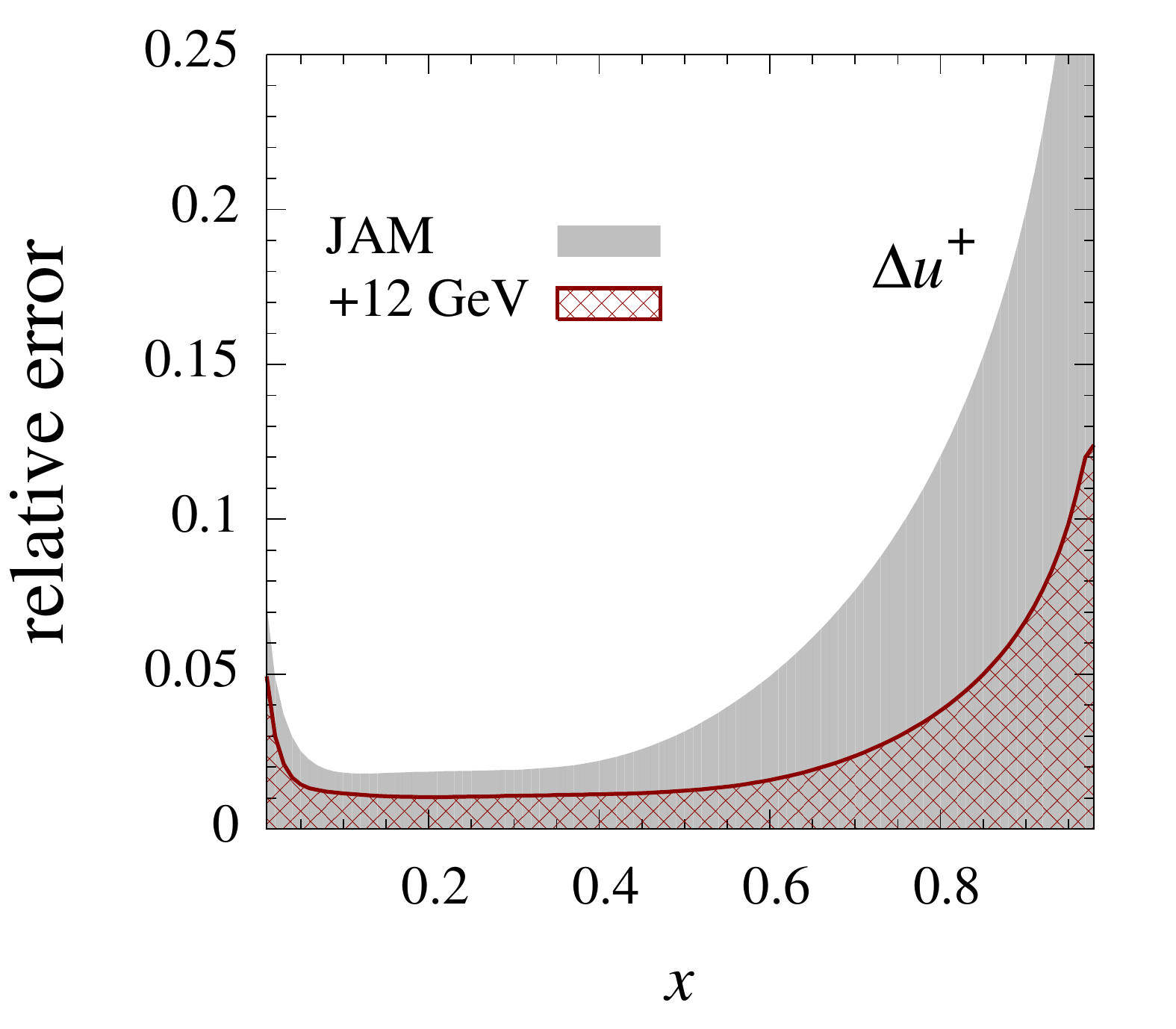}
\includegraphics[width=8cm]{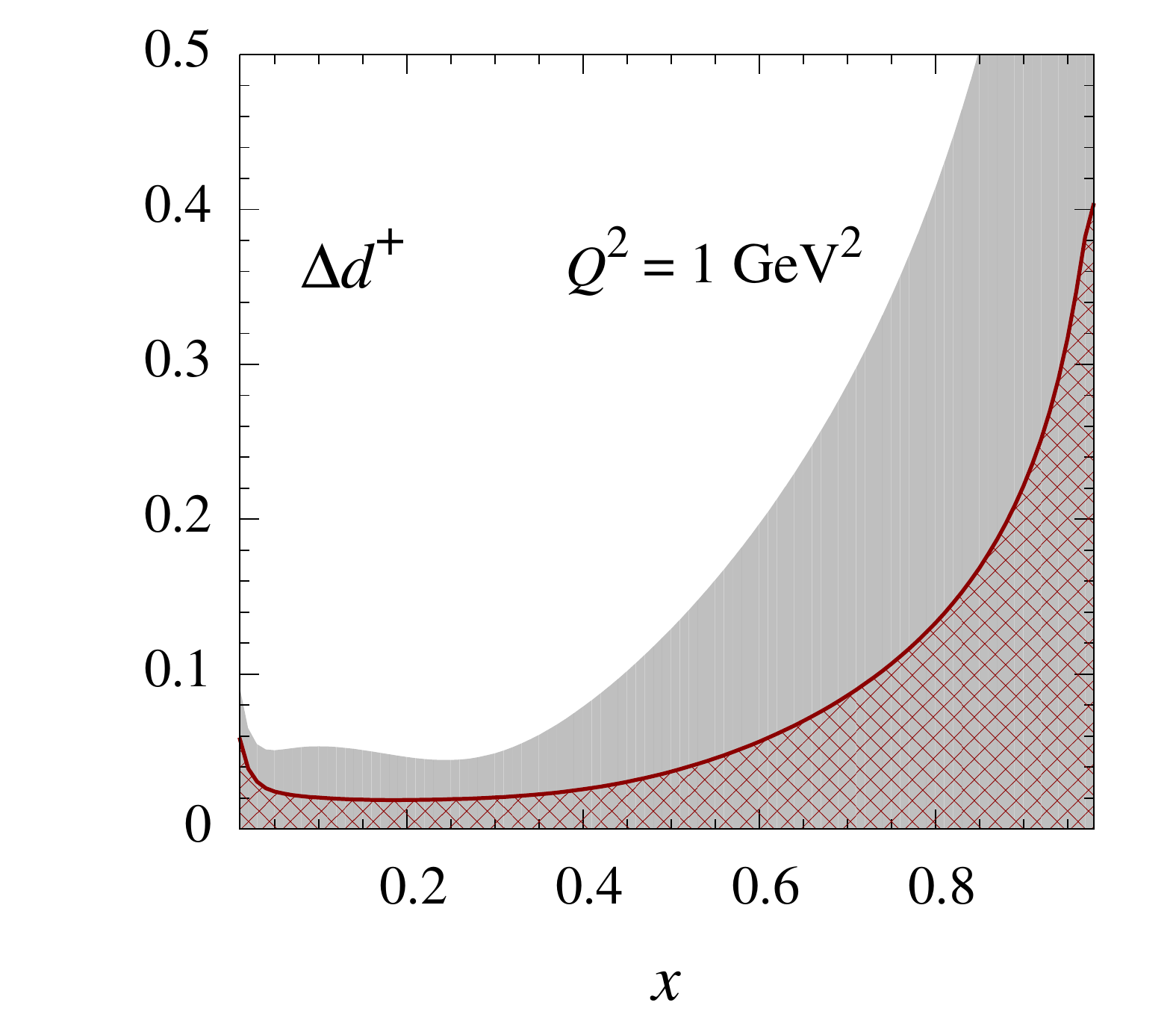}
\caption{Relative error on the $\Delta u^+$ {\bf (left)}
	and $\Delta d^+$ {\bf (right)} PDFs for the JAM fit
	at $Q^2=1$~GeV$^2$ (gray band) and for JAM including
	pseudodata expected from planned Jefferson Lab
	12~GeV experiments \cite{12GeV} (red hashed area).}
\label{fig:12gev}
\end{figure}

To estimate the possible impact of the new Jefferson Lab data we
use the projected statistical and systematic uncertainties for
the proposed experiments at the $x$ and $Q^2$ values where the
asymmetries will be measured \cite{12GeV}.
The pseudodata are generated by randomly distributing the central
values of the points around the JAM fit in Fig.~\ref{fig:data} for
hydrogen, deuterium and $^3$He targets (distributing them around
any of the other fits considered in this analysis would be equally
suitable).  The reduction in the PDF uncertainties, illustrated in
Fig.~\ref{fig:12gev}, is significant, with the relative error on
$\Delta u^+$ and $\Delta d^+$ decreasing by $\sim 70\%$ 
for $x=0.6-0.8$ at the input scale $Q^2=1$~GeV$^2$.

Reductions in the spin-dependent PDF errors such as these,
combined with similarly dramatic reductions expected for
the uncertainty on the unpolarized $d$ quark distribution
(or the $d/u$ ratio) \cite{12GeV_du}, should at the very
least allow one to discriminate between a $\Delta d/d$ ratio
that remains negative, as in simple quark models, and one
that approaches $\sim 1$ in the $x \to 1$ limit, as predicted
by pQCD arguments.
Beyond this there may be additional constraints on the $x \to 1$
behavior of spin-dependent PDFs from an electron-ion collider
\cite{EIC-1, EIC-2, EIC-3}, particularly if the spectator tagging
technique \cite{BONuS} in semi-inclusive DIS from the deuteron
or $^3$He can be extended to polarized beams and targets.

\section{Conclusions}
\label{sec:conc}

The aim of this analysis was to investigate whether existing data
from polarized lepton--nucleon DIS is able to provide any constraints
on the $x \to 1$ behavior of spin-dependent PDFs in the context of a
global QCD analysis.
Using the recent JAM fit as a baseline, we showed that demanding the
polarized to unpolarized PDF ratios $\Delta q^+/q^+$ to approach unity
at $x=1$ results in equally good fits to the available data, even
though the resulting changes to the $\Delta d^+$ PDF are significant
in the intermediate-$x$ region.  With dramatically different behaviors
for the $\Delta d^+/d^+$ ratio allowed for $x \gtrsim 0.3$, this
highlights the critical need for precise data sensitive to the $d$
quark polarization at large $x$ values.

We have investigated the recent suggestion that inclusion of Fock
states in the nucleon wave function with nonzero orbital angular
momentum gives rise to additional contributions to PDFs proportional
to $(1-x)^5 \log^2(1-x)$ \cite{Avakian07} which could play an
important role numerically.  Employing an extension of the typical
functional form used in standard PDF analyses which allows for the
$\log$ dependence, we find that the generalized parametrization is
also able to provide a good description of the existing DIS data,
with or without the $x=1$ constraint.
While there has been a first indication of a rise above unity of
the neutron ($^3$He) polarization asymmetry for $x \gtrsim 0.6$
\cite{E99-117}, the data still generally prefer a negative
$\Delta d^+$ distribution at large $x$ even with the $x=1$
limit built in, although the cross over to positive values
depends on the parametrization chosen (at $\approx 0.95$ for
the JAM+ and $\approx 0.65$ for OAM+).

Further progress on this problem is expected soon with new data
expected from several experiments at the 12~GeV energy upgraded
Jefferson Lab, which aim to measure polarization asymmetries of
protons, deuterons and $^3$He up to $x \sim 0.8$ in DIS kinematics
\cite{12GeV}.
Using the projected statistical and systematic errors from these
experiments, we explored the possible impact on the PDFs and their
uncertainties in this region.
We find reductions in both the $\Delta u^+$ and $\Delta d^+$ PDFs
of up to $\approx 70\%$ for $x \approx 0.6-0.8$ in the JAM fit,
with significant reductions also at smaller $x$ values.
This should considerably narrow the range of possible asymptotic
$x \to 1$ behaviors of the $\Delta q^+/q^+$ ratios, and for the
first time provide critical tests of the various theoretical scenarios
that have been proposed to describe PDFs in the large-$x$ region
\cite{MT96, Feynman72, Close73, Close88, Isgur99, Farrar75, BBS95}.

\acknowledgments

We thank A.~Accardi for helpful discussions, and J.-P.~Chen, S.~Kuhn,
G.~Schnell, B.~Wojtsekhowski and X.~Zheng for communications about
existing and future polarization asymmetry experiments.
This work was supported by the DOE Contract No.~DE-AC05-06OR23177,
under which Jefferson Science Associates, LLC operates Jefferson Lab.

\newpage

\end{document}